\documentclass[journal]{IEEEtran}
\usepackage[T1]{fontenc}
\usepackage[utf8]{inputenc}
\usepackage{amsmath,graphicx,tikz,xcolor,hyperref,booktabs,upquote,csquotes,textcomp,subfig,multirow}
\usepackage{pifont}
\usepackage{xcolor}

\usetikzlibrary{arrows,decorations.pathmorphing,backgrounds,positioning,fadings,shadings,through,fit,petri,calc}

\newcommand{\qt}[1]{\enquote{#1}}
\newcommand{\tab}{\hspace{0.3cm}}

\newcommand{\bq}{\textasciigrave}
\newcommand{\grntxt}[1]{\textcolor{black!50!green}{#1}}
\newcommand{\magtxt}[1]{\textcolor{purple}{#1}}

\newcommand{\blutxt}[1]{\textcolor{blue}{#1}}

\newcommand{\seqtime}{T}
\newcommand{\conctime}{T'}
\newcommand{\confrate}{c}
\newcommand{\ncores}{n}
\newcommand{\speedup}{R}
\newcommand{\prepos}{K}
\newcommand{\ntx}{x}
\newcommand{\abslcc}{L}
\newcommand{\rellcc}{l}

\begin{document}
\title{On Exploiting Transaction Concurrency To Speed Up Blockchains}

\author{
    \IEEEauthorblockN{Dani\"el Reijsbergen and Tien Tuan Anh Dinh} \\[0.1cm]
    \IEEEauthorblockA{Singapore University of Technology and Design}
}

\maketitle

\begin{abstract}
Consensus protocols are currently the bottlenecks that prevent blockchain systems from scaling. However, we argue that
transaction execution is also important to the performance and security of blockchains. In other words, there
are ample opportunities to speed up and further secure blockchains by reducing the cost of transaction
execution. 

Our goal is to understand how much we can speed up blockchains by exploiting transaction concurrency available
in blockchain workloads.   
To this end, we first analyze historical data of seven major public blockchains, namely Bitcoin, Bitcoin Cash, Litecoin,
Dogecoin, Ethereum, Ethereum Classic, and Zilliqa. We consider two metrics for concurrency, namely
the single-transaction conflict rate per block, and the group conflict rate per block. We find that there is more
concurrency in UTXO-based blockchains than in account-based ones, although the amount of concurrency in the
former is lower than expected. Another interesting finding is that some blockchains with larger blocks have more
concurrency than blockchains with smaller blocks. Next, we propose an analytical model for estimating the
transaction execution speed-up given an amount of concurrency. Using results from our empirical
analysis, the model estimates that 6$\times$ speed-ups in Ethereum can be achieved if all available concurrency is exploited.    

\end{abstract}

\section{Introduction}
\label{sec:introduction}
Consensus protocols are currently the fundamental obstacles that prevent blockchain systems from scaling. There is a large
gap between the cost of consensus and the cost of other blockchain layers, in particular the execution and data model
layer~\cite{blockbench}.  Most recent works that seek to improve blockchain performance focus on scaling the consensus
layer, either by designing new protocols~\cite{algorand,hotstuff}, by leveraging sharding~\cite{dang19,
omniledger}, or by weakening security guarantees~\cite{hyperledger}. Despite these efforts, state-of-the-art
blockchains with novel consensus protocols can only achieve a few thousands of transactions per second in
throughput, which is far below what a typical distributed database can do~\cite{blockbench,twin19}. We
argue that it is time to look at other layers of the blockchain for opportunities to increase performance. 

We posit that the execution layer, where blockchain transactions are executed, offers ample
opportunities to improve both the performance and security of blockchains. There are three reasons for that. First,
many modern blockchains employ sharding, which splits the network into small committees that run consensus
protocols independently from the other committees.  Within a small committee, the gap between the cost of consensus
and transaction execution shrinks significantly~\cite{dang19}. In other words, reducing the cost of the
transaction layer can lead to significant performance gains at each committee, which in turn improves
blockchain performance as a whole. Second, some private blockchains such as Hyperledger
Fabric~\cite{hyperledger} abandon classic consensus protocols for other designs that require a trusted third
party service, such as the ordering service discussed in~\cite{hyperledger}. By sacrificing security, these
blockchains shift their bottlenecks away from consensus to other parts of the systems,  one of which being the
execution layer~\cite{bcdl19,sharma19}. Third, the cost of executing transactions negatively affects
blockchains' incentive mechanisms, as captured by the Verifier's Dilemma~\cite{luu15}. As a consequence,
making transaction execution faster strengthens the incentive mechanisms, which in turns strengthens the
overall security. 

We ask the following question: {\em how much can we speed up blockchains by speeding up the execution layer?}
Although a large number of techniques from databases can be employed to speed up transaction execution, we
focus on a single technique: exploiting concurrency to execute multiple transactions in parallel. The fact
that existing blockchains execute transactions in batches (i.e., one batch per block), but within each batch execute transactions only
sequentially, means there could be a large amount of untapped concurrency.  

In this work, we take first steps at answering the above question. We have two goals. The first goal is to
understand the amount of concurrency available in existing blockchains. To this end, we conduct an extensive
empirical analysis of seven public blockchains, namely Bitcoin, Bitcoin Cash, Litecoin, Dogecoin, Ethereum,
Ethereum Classic, and Zilliqa. We choose public blockchains over their private (or permissioned)
alternatives because of their wide adoption and data availability. The selected blockchains cover a large
design space, including state-of-the-art sharding-based systems. We measure concurrency using the
{\em conflict rate per block}: a lower rate means higher concurrency. We compare the seven blockchains
against two variants of this metric: the {\em single-transaction conflict rate}, and the {\em group conflict rate}. Our analysis  
differs significantly from recent work that evaluates concurrency in Ethereum~\cite{saraph2019empirical} in
that our approach is much more lightweight and can extract more concurrency, and that our analysis covers a more
comprehensive dataset that includes more than one blockchain.   

Our second goal is to understand how much execution speed-up can be achieved by exploiting the available
concurrency. To this end, we propose an analytical model for the  computation of the potential speed-up from the conflict rates per
block. An accurate model is not trivial, because it must take into account variables other than conflict rate,
for instance the number of cores per machine, scheduling policies, and sychronization overhead.  

In summary, we make three important contributions. 
\begin{enumerate}
\item We present an extensive data-driven analysis of the amount of concurrency in seven public
blockchains. Our methodology is more lightweight and able to capture more concurrency from more comprehensive
datasets and systems than existing works.  

\item We discuss important findings from the analysis, including: 

$\bullet$ {\em There is more concurrency in UTXO-based blockchains than in account-based ones.}
For example, the rate of single-transaction conflicts in Bitcoin is around $13\%$ whereas in Ethereum it is close to $80\%$.  Although
the difference may seem unsurprising because of the nature of the two data models, a more interesting
observation is that the amount of concurrency in UTXO-based blockchains is lower than expected. 
One extreme example is the Bitcoin block $358624$,\footnote{Hash:\texttt{\scalebox{0.69}{
  0000000000000000162eff5ec874e04dac222a919ca524716eaac575bde4a3da}}.} in which $3217$ out of the total $3264$
  transactions are dependent on each other (i.e., there is no concurrency between them and they must be executed sequentially). 

$\bullet$ {\em In every blockchain, the group conflict rate is lower than the single-transaction conflict rate.} 
Although this is true by definition, the difference is considerable. For example, in Ethereum the former is around $20\%$ whereas the latter is closer to $60\%$ on average. The implication is that there is much
more concurrency to be exploited when transactions are considered in groups as opposed to individually.  


$\bullet$ {\em Blockchains with more transactions per block often have a lower group conflict rate.} For example, on average Ethereum
has an order of magnitude more transactions per block than Ethereum Classic, but its group conflict
rate is much lower than that of the latter, namely $20\%$ compared to $70\%$. The implication is that
blockchains with a higher load potentially have more concurrency, or that blockchains with 
more users have both a higher network load and more concurrency.

\item We present a model that enables the extrapolation of transaction execution speed-ups. Applying the model to the seven
blockchains under consideration, we show potential performance gains of up to 6$\times$.
\end{enumerate}
  
\vspace{0.1in}
The next section presents the relevant background on blockchains and discusses our motivation for speeding up
the execution layer. Section~\ref{sec:methodology} details the methodology of our empirical analysis,
including the metrics and data collection process. Section~\ref{sec:empirical} discusses our findings.
Section~\ref{sec:model} describes the speed-up model. Section~\ref{sec:related} discusses the related work,
before Section~\ref{sec:conclusions} concludes the paper.

\section{Background \& Motivation}
\label{sec:background}
\subsection{Blockchain Systems}
A blockchain system (or blockchain) is a network of nodes that maintain a replicated,
tamper-evident log data structure called a {\em ledger}. The nodes do not necessarily  trust each other.  The ledger is a
sequence of blocks linked together via cryptographic hash pointers. Each block contains multiple transactions
that modify some global state. A blockchain can be examined in four layers: data model layer, consensus
layer, execution layer, and application layer~\cite{blockbench}. The first concerns the storage and nature of the
transactions. The second includes protocols that enable nodes to agree on the ledger. The third concerns how
transactions are executed, and the fourth includes user applications. We refer readers
to~\cite{untangling,blockbench} for a comprehensive discussion of the design space and to \cite{homoliak2019security} for an overview of the potential security pitfalls. 

\vspace{0.1in}
\subsubsection*{Consensus} Consensus protocols are necessary for the security of blockchains because they allow
decentralized, mutually distrustful nodes to agree on the same ledger. Public blockchains, in which any node
can participate, often use variants of Proof-of-Work (PoW) protocols~\cite{btc_origin} which are
computationally intensive. Private (or permissioned) blockchains employ more computionally efficient, classic distributed consensus
protocols such as PBFT~\cite{pbft}. However, these protocols are communication-heavy and do not scale well to
large networks~\cite{dang19}.  

\vspace{0.1in}
\subsubsection*{Data model} The data model determines the nature of the transactions included in the block,
and the operations that can be performed. Most blockchains employ one of the following two
models: UTXO-based and account-based.  

In the UTXO-based model, a transaction takes outputs of other transactions as inputs and creates its own
transaction outputs (or TXOs). Each TXO contains a value. Outputs of one transaction can be taken as input of,
or {\em spent} by, another transaction. A special type of transaction, called {\em coinbase}, has no input UTXOs and produces one output TXO. Nodes keep track of unspent TXOs (or UTXOs). A transaction is valid
if the total value of the output TXOs matches that of the input
TXOs,\footnote{Minus some transaction fees.} and if the input TXOs are in the current UTXO set. 


In the account-based model, a transaction makes modifications to some accounts' states. For example, a payment
transaction updates the state representing the balance in both the sender's and the receiver's account. Executing
a transaction in this model involves the invocation of some computation logics, or smart contracts, that modify the
global state. Together with smart contracts, this model enables blockchain applications that are more complex
and interesting than cryptocurrencies.   


\vspace{0.1in}
\subsubsection*{Smart contract} A smart contract encodes computation over the blockchain states. The contract
is identified by an address, and is triggered by sending a transaction to that address. Smart contracts in different blockchains differ in terms of contract expressiveness,
and in terms of execution runtime. In particular, some blockchains such as Ethereum support Turing-complete
contracts, which allows user to define arbitrary computation, whereas others such as Libra~\cite{libra}
support only a limited set of contracts. Furthermore, some blockchains use specific virtual machines (e.g., the Ethereum Virtual Machine or
EVM) to execute the contract, whereas others rely on general-purposed containers (e.g.  Docker). Given a
 block, existing client software applications execute its transactions sequentially, that is, one transaction at a time and in the
order in which they appear in the block. 

Most blockchains that support smart contracts (particularly Ethereum) allow functions in smart contracts to initiate further calls to other contracts. These interactions do not appear as transactions in the blocks, but can still cause write conflicts. In the rest of this paper, we will refer to these interactions as \emph{internal transactions}. In particular, we define as an internal transaction any interaction between contracts that generates a so-called trace in the {\tt geth} client (which was used to create the Google BigQuery dataset that we use later), and which is not a regular or coinbase transaction. 

\subsection{Public Blockchain Systems}
\begin{table}
\resizebox{0.48\textwidth}{!}{
\begin{tabular}{c|cccc}
 & {\bf Data} &  & {\bf Smart} & {\bf Data} \\
 {\bf Blockchain} &  {\bf model }& {\bf Consensus} & {\bf contracts} & {\bf source} \\ \toprule
Bitcoin & UTXO & PoW & No & BigQuery \\  
Bitcoin Cash & UTXO & PoW & No & BigQuery \\  
Litecoin & UTXO & PoW & No & BigQuery \\  
Dogecoin & UTXO & PoW & No & BigQuery \\  
Ethereum & Account & PoW & Yes & BigQuery \\ 
Ethereum Classic & Account & PoW & Yes & BigQuery \\ 
Zilliqa & Account & PoW+Sharding & Yes & --- \\ 
\end{tabular}}
\caption{Comparison of seven public blockchains. The ``data source column'' indicates whether the blockchain data is
available at sources other than the blockchain client.}
\label{tab:blockchains}

\end{table}

We briefly describe the seven public blockchains that we examine in this paper. These systems cover a large
design space, including state-of-the-art sharding-based blockchains. Table~\ref{tab:blockchains} summarizes
their characteristics.  

\vspace{0.05in}
\subsubsection*{Bitcoin}
Bitcoin is the oldest and most valuable (in terms of market capitalization) blockchain that supports only
cryptocurrency. It has been online since early 2009. Designed as an electronic payment system, Bitcoin enables
participants to exchange the platform's native tokens (which are called \emph{bitcoins}) without the need for
a trusted third party to validate the transactions. It uses the UTXO data model, and Proof-of-Work (PoW) consensus, which requires
nodes to solve hard computational puzzle to create a valid block. This process is called \emph{mining},
and the miners are rewarded for their efforts through the creation of new bitcoins. Bitcoin does not support
smart contracts, but there is a simple scripting language for transactions. 




\vspace{0.05in}
\subsubsection*{Bitcoin Cash}
Bitcoin Cash is a fork of Bitcoin, resulting from a dispute in the Bitcoin community about the maximum block size.
Nodes that follow the Bitcoin Cash protocol accept a larger block size than Bitcoin nodes (8MB instead of
1MB), and are therefore theoretically able to handle a higher throughput. Bitcoin Cash has the same chain of blocks 
as Bitcoin until July 2017.

\vspace{0.05in}
\subsubsection*{Litecoin}
Litecoin is a Bitcoin spin-off that is mostly based on the original Bitcoin, except for some minor
modifications including a higher block frequency. Litecoin was launched in October 2011.

\vspace{0.05in}
\subsubsection*{Dogecoin}
Dogecoin is designed as a light-hearted cryptocurrency blockchain based on the now-defunct Luckycoin, which itself is a
spin-off of Litecoin. Dogecoin has a higher block frequency than Litecoin, but is otherwise similar to Bitcoin
and Litecoin. It has been online since December 2013.

\vspace{0.05in}
\subsubsection*{Ethereum}
Ethereum is the first blockchain platform that supports Turing-complete smart contracts. It uses a memory-hard
variant of PoW and an account-based data model. Smart contracts are identified via an address, in the same way
as a regular account.  Ethereum miners and other validating nodes execute the transactions in the blocks in
the Ethereum Virtual Machine (EVM). Each operation in the EVM incurs a cost called \emph{gas} that is
proportional to its computational cost. The gas system prevents denial-of-service or bugs caused by infinite loops
and overly costly operations. Ethereum has been online since July 2015.

\vspace{0.05in}
\subsubsection*{Ethereum Classic}
Ethereum Classic is a fork of Ethereum that occurred in July 2016 following a dispute over governance after
the attack on the DAO contract~\cite{atzei2017survey}. Ethereum Classic is currently still highly similar to Ethereum. 

\vspace{0.05in}
\subsubsection*{Zilliqa}
Zilliqa~\cite{zilliqa} is one of the first blockchains that use sharding to increase throughputs. In particular, it employs
{\em network sharding} which assigns nodes to small committees such that each
committee maintains the complete global state. Zilliqa adopts the account-based data model and supports
Turing-complete smart contracts. Its scalability comes from the fact that transactions are processed
independently at different committees that are selected based on the senders' addresses. The system uses a
combination of PoW and classic consensus protocols. In particular, nodes run PoW to determine their
committees, and a variant of PBFT~\cite{pbft} to ensure security at local committees. A major limitation of
Zilliqa is that it does not support cross-shard transactions --- ones that touch multiple committees. In
addition, it needs to wait for state synchronization between committees before transactions are confirmed.
Recent blockchains have addressed these limitations~\cite{dang19,omniledger}, but we consider Zilliqa in this
work because it is the only sharding-based public blockchain that is running live and that has a
considerable amount of traffic.

\subsection{Why Improve the Execution Layer?}
We discuss three reasons why it is beneficial to make the execution layer more efficient. First, as shown
in~\cite{dang19}, for small networks, the cost of transaction execution is comparable to, if not greater than,
the cost of consensus. In particular, for a 7-node private blockchain, the average execution time per block is
$250ms$ while the average consensus time is $20ms$. For a 31-node network, the average cost for execution and
consensus are similar at around $250ms$. In other words, the execution layer has a large contribution to the
overall cost, therefore making it efficient will lead to significant performance gains. It might be
countered that this observation does not hold in practice because blockchain networks are large. However, we
note that most modern blockchains use sharding to break up the network into much smaller committees (or
sub-networks), which means that within each sub-network the cost of the execution layer remains significant. 

Second, some blockchains sacrifice the security of the consensus protocols for performance, by abandoning PoW and
other classic Byzantine fault-tolerant protocols. For example, Hyperledger Fabric employs a Kafka cluster,
which does not tolerate Byzantine failures, to achieve transaction order. As shown in~\cite{bcdl19}, such
designs can shift the bottleneck away from consensus. For example, in Hyperledger Fabric, the bottleneck is the
endorsing phase which executes (simulates) transactions before sending them to Kafka. Since the execution layer
becomes a likely bottleneck in these blockchains, reducing its cost leads to significant performance gains.   

Finally, the cost of transaction execution negatively affects the security of public blockchains, as captured
by the Verifier's Dilemma~\cite{luu15}. In particular, a rational node has considerable incentive to skip the
transaction execution, and to spend all of its resources on consensus (for instance, to mine new blocks).
But without a large number of nodes executing the same transactions, the overall security becomes lower because
invalid and malicious transactions can slip through and get recorded in the ledger. In other words, reducing
the cost of transaction execution helps to strengthen security, because rational nodes have less incentive to
skip transaction execution.

\section{Methodology}
\label{sec:methodology}
\begin{figure*}
\centering
\subfloat[][Ethereum block 1000007.]{
 \includegraphics[scale=0.9225]{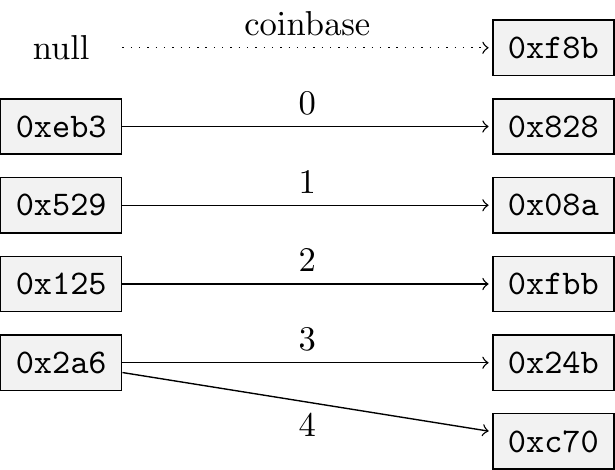}
 \label{fig:block_1000007}
}
\hspace{2.5cm}
\subfloat[][Ethereum block 1000124.]{
 \includegraphics[scale=0.9225]{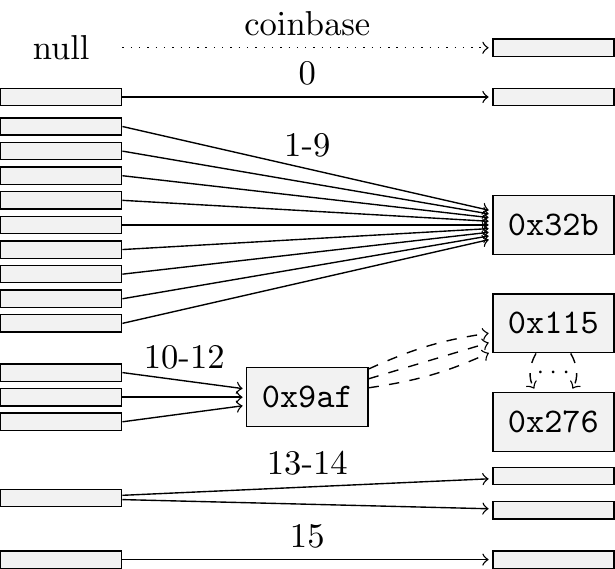}
 \label{fig:block_1000124}
} 
\caption{Examples of transaction dependency graphs, for Ethereum blocks 1000007
and 1000124. Solid, dotted and dashed lines represent regular, coinbase and internal transactions
respectively.} \label{fig:ethereum_tdgs}
\end{figure*}

In this section we present the methodology behind our empirical study of the seven blockchain platforms. We
begin in Section~\ref{sec:metrics} with a discussion of the metrics that we seek to compute. We discuss our
data sources in Section~\ref{sec:data}, and the queries for obtaining the metrics of interest from the online
datasets in Section~\ref{sec:sql_queries}.
 
\subsection{Concurrency Metrics}
\label{sec:metrics}
We quantify concurrency per block via \emph{conflict rates}: the lower the conflict rate, the higher the
amount of concurrency.  In the following, we first define conflicts in a model called the {\em transaction
dependency graph}. We then derive two metrics for concurrency, namely single-transaction concurrency, and
group concurrency. 

\vspace{0.1in}
\subsubsection{Transaction Dependency Graph (TDG)} 
Each block is modelled as a graph $(N,E)$ where $N$ denotes the set of nodes and $E$ the set of edges. The nature of nodes
and edges depends on the data model. 
\begin{itemize}
\item UTXO-based models: each node is a transaction
in the block. An edge exists from $a$ to $b$, i.e., $(a,b) \in E$, if a TXO is created in node $a$ and spent in node $b$.  
\item Account-based models: each node is an address that is referenced by a transaction in the block. An edge
exists from $a$ to $b$ if there is a (possibly internal) transaction in which $a$ is the sender
and $b$ is the receiver.  
\end{itemize}



For both data models, we ignore the coinbase transctions for simplicity.
Any two edges in TDG that share an endpoint are said to be \emph{connected}. A \emph{path} is a
sequence of connected edges. Two nodes $a,b \in N$ are said to be connected if a path exists such that $a$ is
the first endpoint and $b$ the final endpoint. A set of nodes $C \subseteq N$ forms a a \emph{connected
component} of size $|C|$ if all pairs of nodes in $C$ are connected, yet no single other node in $N \backslash
C$ exists that is connected to any node in $N$.   

\vspace{0.1in}
\subsubsection{Transaction conflict} In the UTXO-based model, we say that a transaction $t \in N$
\emph{conflicts} with another transaction \mbox{$u \in N$} if $t$ and $u$ are part of the same connected component.
In the account-based model, we say that a transaction $(a,b)$ conflicts with another transaction $(c,d)$ if
their endpoints are part of the same connected component. We say that a transaction is \emph{conflicted} if it
conflicts with any other transaction. 

\vspace{0.1in}
\subsubsection{Concurrency metrics}
The following two metrics capture the amount of concurrency in a block. 
\begin{itemize}
\item {\bf Single-transaction conflict rate.} We define the single-transaction conflict rate as the ratio between the number of
conflicted transactions and the total number of transactions within the block. 

\item {\bf Group conflict rate.} Let the {\em LCC size} be the size of the largest connected component (LCC) in a block. We
define the group conflict rate as the relative LCC size, that is, the ratio between the LCC size and the total number of
transactions in the block. 
\end{itemize}

When we display the evolution of these metrics for the historical datasets, we will always \emph{weight} these metrics by the block size (or gas cost). The reason is that larger blocks contribute more to the blockchain's total execution cost than smaller ones.


\vspace{0.1in}
\subsubsection{Examples}
Figure~\ref{fig:ethereum_tdgs} shows two examples of TDGs for account-based models, namely for Ethereum blocks
$1000007$ and $1000124$. The first block (Figure~\ref{fig:block_1000007}) contains 5 transactions and
one coinbase transaction. The number above a transaction indicates its index in the block's
transaction list. In this block, transactions $3$ and $4$ are conflicting because their endpoints are part of
the same connected component. According to {\tt etherscan.io}, the {\tt 0x2a6}...  address that causes the
conflict belongs to the DwarfPool mining pool. If we ignore the coinbase transaction, then the
block of Figure~\ref{fig:block_1000007} has 5 transactions and 4 connected components, namely $3$ of size $1$
and $1$ of size $2$.  Two of its transactions are conflicted, so its single-transaction
conflict rate is $40\%$, and the group conflict rate is also $40\%$. 

The second block (Figure~\ref{fig:block_1000124}) contains 15 regular transactions, one coinbase transaction,
and 18 internal transactions.  Transactions 1-9 all send funds to the same address, which according to {\tt
etherscan.io} is owned by the Poloniex cryptocurrency exchange. Transactions 10-12 were sent to a smart
contract (which is unverified but which received $73,369$ transactions between January and March 2016). This
contract in turn makes one call to another unverified contract, which then contacts the contract at {\tt
0x276}..., which is a verified contract called ElcoinDb associated with the \qt{ElCoin} ERC20 token.
Transactions 13 and 14 are sent by the same address, which belongs to DwarfPool. The block contains 5 connected components. Furthermore, $14$ out of its $16$
transactions are conflicted, so its single-transaction conflict rate is $87.5\%$, but the group conflict rate
is lower at $56.25\%$. 

\vspace{0.1in}
\subsubsection{Discussion}
We remark that our definition of conflicting transactions is different to that in~\cite{saraph2019empirical}.
In particular, the latter defines conflict as accessing the same storage location, which means that two
transactions sent to the same addressed may not be considered conflicted if they invoke different methods
and access different states. Additionally, it is not entirely clear from the discussion of their algorithm whether, during a situation where several transactions access the same memory location, the first transaction that does so is also considered as conflicting, or if only the later ones are placed in the conflicting transaction `bin'. Finally, the analysis in~\cite{saraph2019empirical} does not consider pure
payment transactions, which leads to fewer transactions per block. Because of these, single-transaction
conflict rates reported in~\cite{saraph2019empirical} are lower than in ours, indicating higher concurrency.
However, as shown in Section~\ref{sec:empirical}, by using group conflict instead of single-transaction conflict we are able to extract more concurrency.  


\subsection{Data Collection}
\label{sec:data}
We collect real data from seven public blockchains. Most of the datasets are available on Google
BigQuery,\footnote{\url{https://cloud.google.com/bigquery/}} which also supports large-scale query processing. 
We leverage this service for six out of the seven
blockchains.\footnote{\url{https://cloud.google.com/blog/products/data-analytics/introducing-six-new-cryptocurrencies-in-bigquery-public-datasets-and-how-to-analyze-them}}
Most of these datasets follow a similar format. The datasets for UTXO-based systems follow the schema of the
Bitcoin dataset, and Ethereum Classic follows the schema of the original Ethereum dataset. They are queried
using SQL and user-defined functions (UDFs) written in JavaScript, as described in the following section.

Zilliqa is not included in Google BigQuery public datasets, thus we implemented a lightweight client for
downloading the data from Zilliqa's mainnet. The client is written in Python and uses Zilliqa's Python SDK for
querying the blockchain. It works in two phases. In the first phase, it downloads all transaction hashses using
{\tt GetTransactionsForTxBlock} method. In the second phase, it downloads details for every transaction
obtained from the first phase, using the {\tt GetTransaction} method. Although the SDK throughputs are low (namely about
$4$ request per second), the collection of the entire Zilliqa blockchain is fast because there are only $360K$
blocks and $2.2M$ transactions. 


\subsection{SQL Queries}
\label{sec:sql_queries}

Since the six Google BigQuery datasets follow only two schemas, it is sufficient to construct two queries
to process the data. Most of the computationally expensive parts are done using Javascript UDFs.
An example query for the UTXO-based systems is shown in Figure~\ref{fig:bitcoin_sql_query}. For
each block, the query creates two equally-sized arrays, {\tt inputs\_merged.txs} and {\tt
inputs\_merged.spent\_txs}. The $i$th element in the former array is the hash of the transaction that creates
the $i$th input TXO, and the $i$th element in the latter is the hash of the transaction that spends the
$i$th input TXO. The TDG is then constructed as follows: since each transaction is a node,
every pair in the two arrays defines an edge. 
\begin{figure}
\centering
\framebox[0.49\textwidth]{
\hspace{0.05cm}
\begin{minipage}[t]{0.49\textwidth}
\begin{flushleft}
\footnotesize{\texttt{\magtxt{SELECT}  \\
\tab block\_number,  \\
\tab blocks.block\_data[OFFSET(\grntxt{0})] \magtxt{AS} num\_transactions,  \\
\tab blocks.block\_data[OFFSET(\grntxt{1})] \magtxt{AS} num\_conflict\_txs,  \\
\tab blocks.block\_data[OFFSET(\grntxt{2})] \magtxt{AS} max\_lcc\_size  \\
\magtxt{FROM} ( \\
\tab \magtxt{SELECT}  block\_number,  \\
\tab \tab process\_graph(ARRAY\_AGG(inputs\_merged.txs), \\ 
\tab \tab \tab ARRAY\_AGG(inputs\_merged.spent\_txs))  \\
\tab \magtxt{AS} block\_data  \\
\tab \magtxt{FROM} ( \\
\tab \tab \magtxt{SELECT}  txs.block\_number \magtxt{AS} block\_number,  \\
\tab \tab \tab inputs.spent\_transaction\_hash \magtxt{AS} spent\_txs,  \\
\tab \tab \tab txs.hash \magtxt{AS} txs  \magtxt{FROM}  \\
{\bq}bigquery-public-data.crypto\_bitcoin.transactions{\bq} \\
\tab \tab \magtxt{AS} txs,  UNNEST(inputs) \magtxt{AS} inputs \\
\tab ) \magtxt{AS} inputs\_merged \\
\tab \magtxt{GROUP} \magtxt{BY} block\_number \magtxt{ORDER} \magtxt{BY} block\_number \\
) \magtxt{AS} blocks \\
}}
\end{flushleft}
\end{minipage}
}
\caption{An SQL query for the Bitcoin-like datasets. }
\label{fig:bitcoin_sql_query}
\end{figure}

\begin{figure}
\centering
\framebox[0.45\textwidth]{
\hspace{0.2cm}
\begin{minipage}[t]{0.45\textwidth}
\begin{flushleft}
\footnotesize{\texttt{var \blutxt{ccs} = []; \\
for(\magtxt{var} \blutxt{i}=\grntxt{0};i<txs.length;i++) \{ \\
\tab \magtxt{if}(visitedMap[txs[i]] == \grntxt{0}) \{ \\
\tab \tab \magtxt{var} \blutxt{cc} = [txs[i]]; \\
\tab \tab visitedMap[txs[i]] = \grntxt{1}; \\
\tab \tab \magtxt{var} \blutxt{neighbors} = \magtxt{new} Set(); \\
\tab \tab \magtxt{for}(\magtxt{let} \blutxt{nb} \magtxt{of} nbMap[txs[i]]) \{ \\
\tab \tab \tab neighbors.add(nb); \\
\tab \tab \} \\
\tab \tab \magtxt{while}(neighbors.size > \grntxt{0}) \{ \\
\tab \tab \tab \magtxt{var} \blutxt{newNeighbors} = \magtxt{new} Set(); \\
\tab \tab \tab \magtxt{for}(\magtxt{let} \blutxt{nb} \magtxt{of} neighbors) \{ \\
\tab \tab \tab \tab cc.push(nb); \\
\tab \tab \tab \tab visitedMap[nb] = \grntxt{1}; \\
\tab \tab \tab \tab \magtxt{for}(\magtxt{let} \blutxt{nnb} \magtxt{of} nbMap[nb]) \{ \\
\tab \tab \tab \tab \tab \magtxt{if}(visitedMap[nnb] == \grntxt{0}) \{ \\
\tab \tab \tab \tab \tab \tab newNeighbors.add(nnb); \\
\tab \tab \tab \} \} \} \\
\tab \tab \tab neighbors = newNeighbors; \\
\tab \tab \} \\
\tab \tab ccs.push(cc); \\
\} \} \\
}}
\end{flushleft}
\end{minipage}
}
\caption{An implementation of breadth-first search. }
\label{fig:bitcoin_main_loop}
\end{figure}

The {\tt process\_graph} computes the metrics of interest. Its main job is to first create the TDG, and then
to determine the connected components using breadth-first search. We use three associative arrays as helpers.
The first is {\tt nbMap}, which maps each transaction to its neighbors in the graph, i.e., those
nodes with which it shares an edge. The second is {\tt inBlockMap} which tracks the block the transaction
appears  in. 
The third is {\tt visitedMap} which tracks whether a
transaction has been visited during breadth-first search. 
%
The core of the algorithm is shown in Figure~\ref{fig:bitcoin_main_loop}. The result, {\tt ccs}, is an array
of arrays, where each element of the main array contains all the hashes of a connected component. 
Once the algorithm has finished, the number of unconflicted transactions equals the number of elements of
{\tt ccs} with length 1, and the LCC can be obtained by finding the element of {\tt ccs} with the largest
size.

The query for Ethereum is similar, and is only different in terms of how the nodes and edges are
defined, and requires one more step where the connected components for the addresses are mapped to the
transactions. Finally, for Ethereum we also pass a list of transaction gas costs to the UDF, in order to
collect gas-weighted metrics as we discuss in the next section. For Zilliqa, we first exported the data to a CSV file, after which we used a similar procedure written in Java to obtain the TDGs.

\section{Empirical Analysis}
\label{sec:empirical}
In this section we present the main findings from the empirical analysis of seven blockchain datasets. In
Section~\ref{sec:data_model} we compare the concurrency in UTXO-based versus account-based blockchains. In
Section~\ref{sec:concurrency_type} we examine the differences between the two main concurrency metrics, namely
the single-transaction and group conflict rates. Finally, in Section~\ref{sec:block_size} we discuss the
relationship between concurrency and average block sizes.  

We use the SQL queries in the previous section to compute the two metrics for every block in the history of
the seven blockchains. The figures are generated by dividing these histories into fixed-size buckets for which
we compute weighted averages. The number of buckets ranges from $20$ to $200$. To improve the accuracy of our
results, we weight each block either by the number of transactions or by its gas
consumption. In particular, if there is a high variance between the blocks in terms of the number of
transactions or the amount of gas, then the blocks having more transactions or consuming more should
be weighted more heavily, because they have a greater impact on the total execution time. 

\subsection{UTXO-Based vs. Account-Based Models}
\label{sec:data_model}

\begin{figure*}
    \centering
    \subfloat[][Number of regular/total transactions per block]{
    	\includegraphics[width=0.33\textwidth]{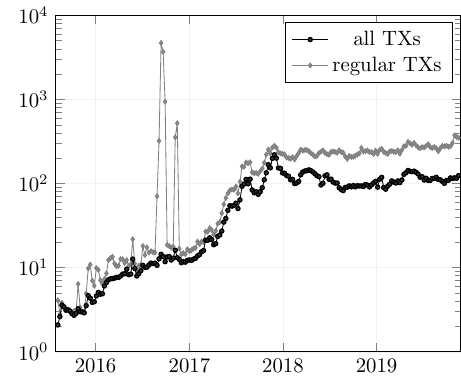}
		\label{fig:ethereumNTx.pdf}
	}
    \subfloat[][Single-transaction conflict rate (weighted)]{
    	\includegraphics[width=0.33\textwidth]{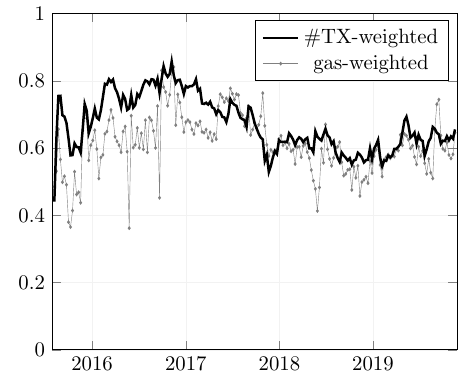}
		\label{fig:ethereumTxConflRatioWeighted.pdf}
	}
	\subfloat[][Group conflict rate (weighted)]{
    	\includegraphics[width=0.33\textwidth]{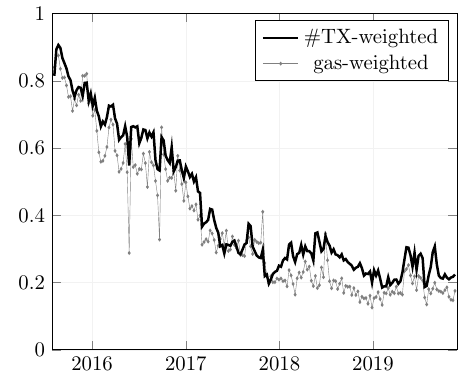}
		\label{fig:ethereumLccSizeRatioWeighted.pdf}
	}
    \caption{Ethereum: evolution over time of the transaction load and the conflict rates.}
    \label{fig:ethereum}
\end{figure*}

\begin{figure*}
    \centering
    \subfloat[][Number of transactions/traces per block]{
    	\includegraphics[width=0.33\textwidth]{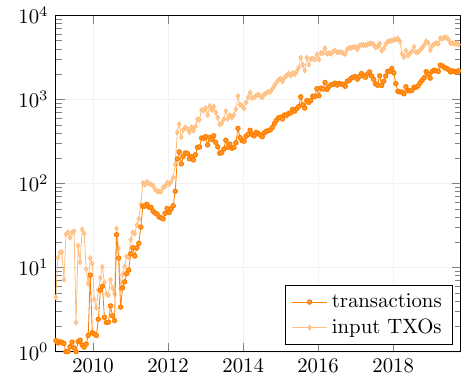}
		\label{fig:bitcoinNTx.pdf}
	}
    \subfloat[][Single-transaction conflict rate (weighted)]{
    	\includegraphics[width=0.33\textwidth]{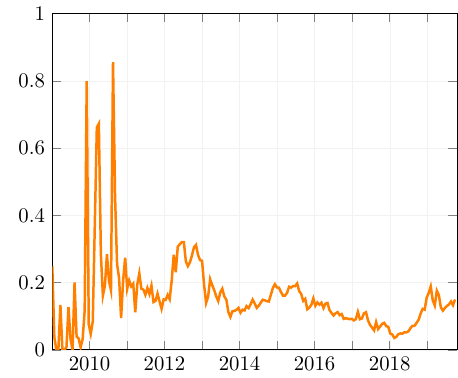}
		\label{fig:bitcoinTxConflRatio.pdf}
	}
	\subfloat[][Group conflict rate (weighted)]{
    	\includegraphics[width=0.33\textwidth]{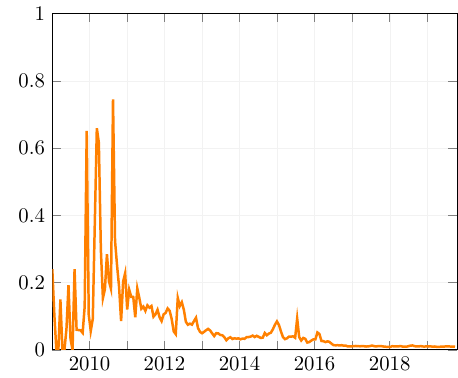}
		\label{fig:bitcoinLCCRatio.pdf}
	}
    \caption{Bitcoin: evolution over time of the transaction load and the conflict rates.}
    \label{fig:bitcoin}
\end{figure*}

We begin our comparison of the two data models with a detailed comparison of the two main cryptocurrency
blockchains: Bitcoin and Ethereum. The results for Ethereum are shown in Figure~\ref{fig:ethereum}.
Figure~\ref{fig:ethereumNTx.pdf} shows that the per-block average number of regular transactions
is around 100 per block, and $300$ if we include internal transactions. The sharp peaks in the number of
internal transactions in the second half of 2017 are probably due to denial-of-service attacks that exploited
EVM instructions that were underpriced~\cite{atzei2017survey}. The single-transaction
conflict rates shown in Figure~\ref{fig:ethereumTxConflRatioWeighted.pdf} are weighted by transaction count
(thick line) and gas (thin line), respectively. We observe that the transaction-weighted conflict ratio is
high, starting around $80\%$ in 2016 and 2017 before decreasing to around $60\%$. By contrast, the
gas-weighted conflict ratio is roughly $60\%$ since Ethereum's early days. One possible reason for this difference is that certain
transactions with a very high gas cost (particularly contract creations) are less likely to be conflicting,
since it is unusual for a single user to create more than one contract per block due to the high cost of doing
so.  Figure~\ref{fig:ethereumLccSizeRatioWeighted.pdf} shows the group conflict rate, which had a period of
decrease until early 2018, and has been stable around $20\%$ since then.

\begin{figure*}
\centering
\includegraphics[scale=0.9225]{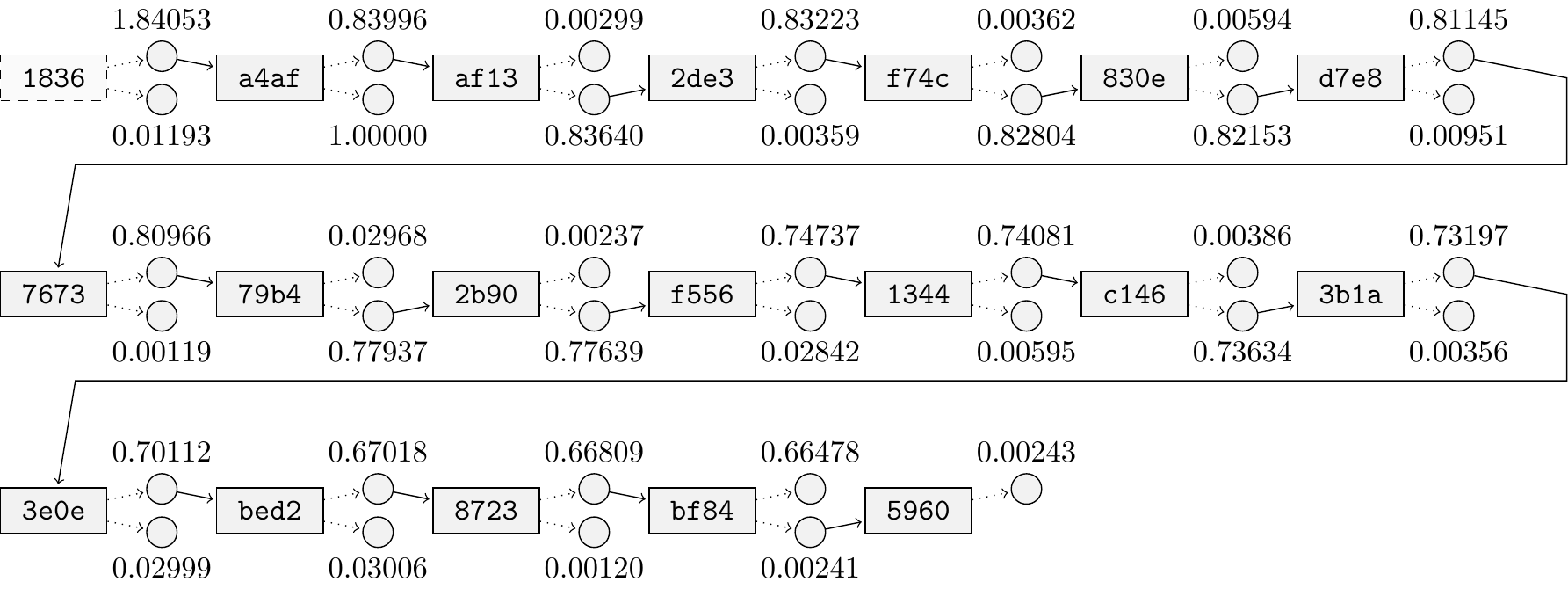}
\caption{Example of a transaction sequence in Bitcoin that occurs in the block $B$ at height 500,000.
Rectangles indicate transactions, and contain the first four hexadecimal digits of its hash. Solid rectangles
occur within $B$, and dashed rectangle in other blocks. Circles indicate TXOs (the values are displayed in
bitcoins up to five decimals of accuracy). Dotted lines connect transactions to their output TXOs, and a solid
line indicates that a TXO was used as an input TXO for another transaction. The sequence start with the
transaction {\tt \scriptsize 1836b68048373543a5e3557c5b192a92eae07ff7cf0588fceff332bba4e6214f}, which occurs
in the block at height 499975. It has two outputs, of which the one with value $1.84052715$ in spent in a
transaction in $B$. This transaction again has two outputs, of which one is again spent in one of $B$'s
transactions. This pattern continues, resulting in a sequence of $18$ transactions within $B$. The
transactions within this sequence must be executed sequentially.} \label{fig:sequence}
\end{figure*}

Figure~\ref{fig:bitcoin} shows the same graphs for Bitcoin, in which the conflict rates are weighted by the
number of transactions. We observe that the average number of transactions per block is currently
over 2000, which is greater than for Ethereum. The average number of input TXOs per block is around 4000.
However, the single-transaction conflict rate for Bitcoin is currently much lower than for Ethereum, namely
roughly $15\%$ compared to $60\%$. The group conflict rate is even lower, namely around $1\%$. This is to be
expected: unlike accounts, which can send or receive transactions many times, TXOs can only be created or
spent once. The only source of conflict in the UTXO-based model is when a TXO is created and spent within the
same block. In fact, the frequency with which this occurs is surprisingly high, and may be due to mining
pools, centralized cryptocurrency exchanges, or because of higher-level protocols being executed on top of
Bitcoin via its scripting language. An example of a long sequence of Bitcoin transactions creating and
spending each other's TXOs is shown in Figure~\ref{fig:sequence}. This example is from the Bitcoin block
$500,000$. We observe that such sequences on average only form a relatively small part of the block.

\begin{figure*}
    \centering
    \subfloat[][Single-transaction conflict rate (weighted)]{
    	\includegraphics[width=0.24\textwidth]{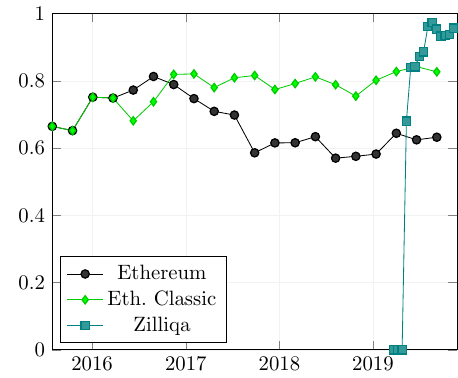}
	}
	\subfloat[][Single-transaction conflict rate (weighted)]{
    	\includegraphics[width=0.24\textwidth]{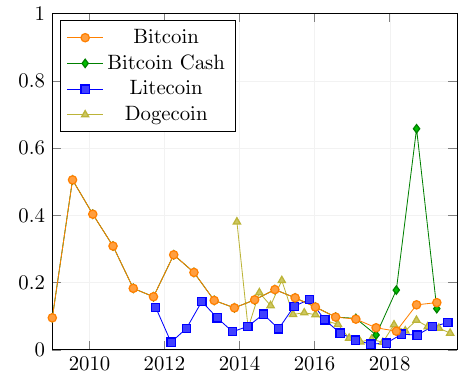}
	}
	\subfloat[][Group conflict rate (weighted)]{
    	\includegraphics[width=0.24\textwidth]{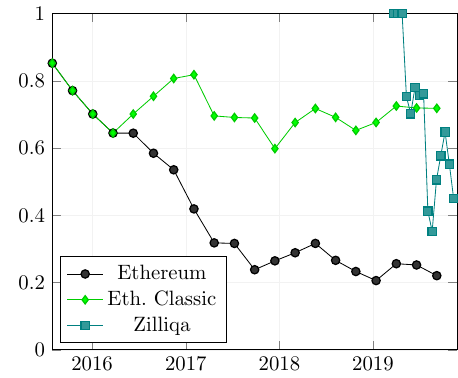}
	}
	\subfloat[][Group conflict rate (weighted)]{
    	\includegraphics[width=0.24\textwidth]{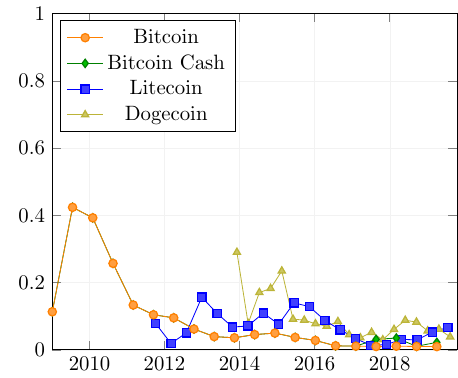}
	}
    \caption{The evolution over time of the conflict rates for all 7 blockchains, grouped by data model.}
    \label{fig:all_coarse}
\end{figure*}

Figure~\ref{fig:all_coarse} compares the conflict rates for all seven blockchains, with the
UTXO-based and the account-based ones grouped separately. The same patterns can be observed, namely that
all conflict rates are considerably lower for the UTXO-based blockchains than for the account-based ones.
However, we note that the account-based blockchains tend to support a wider and more computationally expensive
functionality (i.e., smart contracts) than the UTXO-based ones (which mostly support cryptocurrencies with
very limited scripting support). As the consequence, the higher concurrency on the latter may not translate to
higher speed-ups in absolute terms than the former. Finally, we attribute the high conflict rates in Zilliqa to its workload
characteristics, since the sharding design does not introduce any properties that may explain such a high
rate.

\subsection{Transaction vs. Group Concurrency}
\label{sec:concurrency_type}

Another conclusion that can be drawn from Figure~\ref{fig:all_coarse} is that the group conflict rate is
significantly lower than the single-transaction conflict rate. This is to be expected: after all, unless all
transaction are mutually independent, then all transactions in the largest connected component are necessarily
conflicted, so the single-transaction conflict must always be at least as high as the group conflict rate.
However, the difference is large, for example the single-transaction conflict rate for Ethereum is around $60\%$
and the group conflict rate is around $20\%$. This suggests that techniques that exploit group concurrency have
much more speed-up potential than ones that only focus on individual transactions.

\subsection{Small vs. Big Blocks}
\label{sec:block_size}

\begin{figure*}
    \centering
    \subfloat[][Number of transactions per block]{
    	\includegraphics[width=0.33\textwidth]{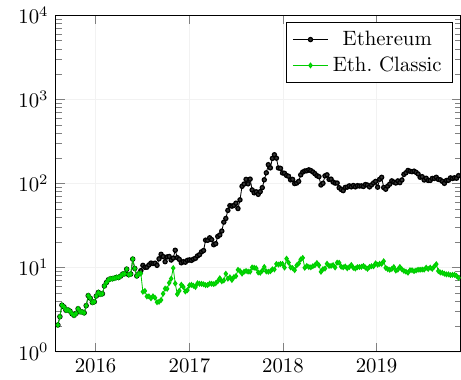}
	}
    \subfloat[][Single-transaction conflict rate (weighted)]{
    	\includegraphics[width=0.33\textwidth]{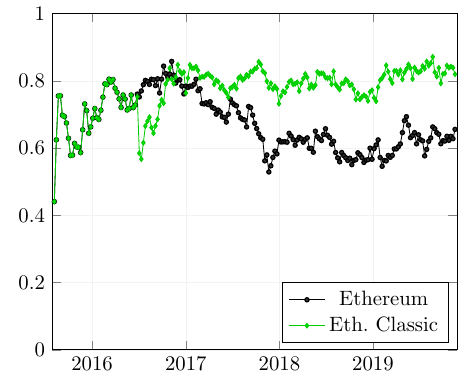}
	}
	\subfloat[][Group conflict rate (weighted)]{
    	\includegraphics[width=0.33\textwidth]{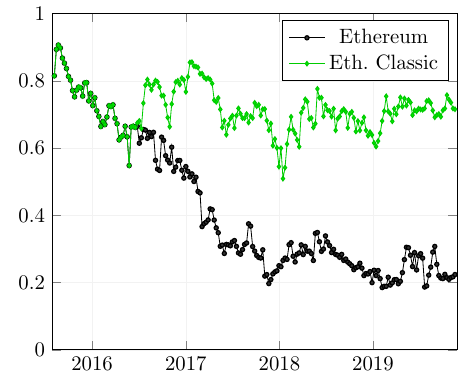}
	}
    \caption{Detailed comparison of Ethereum and Ethereum Classic.}
    \label{fig:ethereum_classic}
\end{figure*}

We examine whether the average number of transactions has an impact on the conflict
rates. In particular, we focus on the difference between Ethereum and Ethereum Classic. A fine-grained
comparison of the two is shown in Figure~\ref{fig:ethereum_classic}. As we can see, Ethereum
Classic has an order of magnitude fewer transactions than Ethereum since early 2018. However, both the
single-transaction and group conflict rates are higher in  Ethereum Classic than in Ethereum, in the latter
case considerably so. This may be surprising, especially for the single-transaction case: if the size of the
user base is similar, then a higher number of transactions per block means that the probability that two
transactions conflict is higher. However, since this does not appear to be the case, this must mean that the
user base for Ethereum Classic is relatively smaller than Ethereum's.

\begin{figure*}
    \centering
    \subfloat[][Number of transactions/traces per block]{
    	\includegraphics[width=0.33\textwidth]{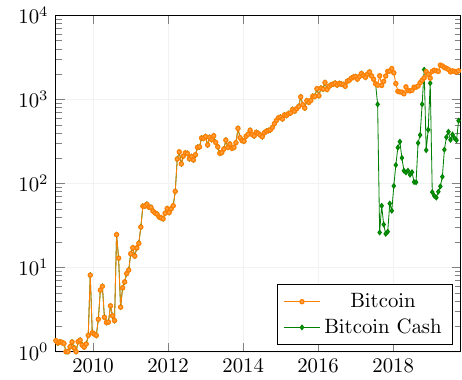}
		\label{fig:btc_bth_nTx.pdf}
	}
    \subfloat[][Conflict ratio per block]{
    	\includegraphics[width=0.33\textwidth]{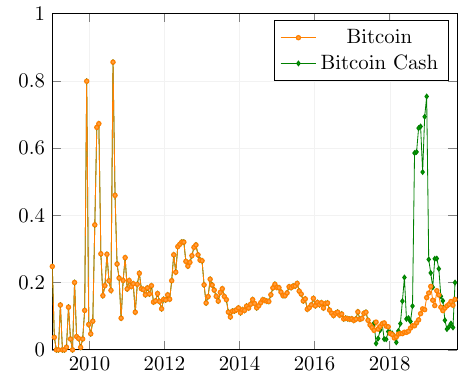}
	}
	\subfloat[][Absolute LCC size per block]{
    	\includegraphics[width=0.33\textwidth]{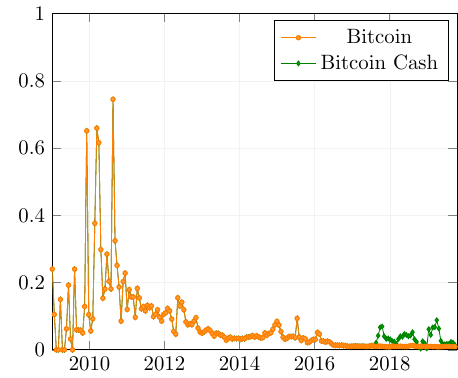}
	}
    \caption{Detailed comparison of Bitcoin and Bitcoin Cash}
    \label{fig:bitcoin_cash}
\end{figure*}

The same comparison for Bitcoin and Bitcoin Cash is shown in Figure~\ref{fig:bitcoin_cash}. We observe from
Figure~\ref{fig:btc_bth_nTx.pdf} that for most of its history, Bitcoin Cash has fewer transactions than
Bitcoin, at times more than an order of magnitude fewer. This is remarkable, as the stated reason for the
creation of Bitcoin Cash was to enable bigger blocks and therefore a larger maximum number of transactions.
Despite the smaller number of transactions, both conflict rates were higher for Bitcoin Cash than for Bitcoin.
Again, this is an evidence that the user base of Bitcoin Cash is smaller, with the big exchanges producing a
larger share of the traffic.

\section{Execution Speed-Up Model}
\label{sec:model}
In this section we discuss how the two concurrency metrics -- the single-transaction and group
conflict rates -- can be used to predict the potential speed-ups of transaction execution in a block. Two
transactions cannot be executed concurrently if they access the same memory, which in UTXO-based models
means that they access the same elements of the UTXO set, and in account-based models that they access the same account and/or
state variables. We note that the TDG contains all necessary information about the potential conflicts: if two
transactions are not part of the same connected component, then at no point do they (or internal transactions
resulting from them) conflict. This informs the two approaches below that approximate execution speed-ups in
a model where transactions in a block have the same execution time. We begin in
Section~\ref{sec:cr_speedup} by describing the technique based on~\cite{saraph2019empirical}, and highlight
that our contributions are the closed-form expressions for the speed-up potentials. In
Section~\ref{sec:lcc_speedup} we describe a technique based on group concurrency. Finally, we present an
empirical evaluation of the potential speed-ups based on historical datasets for Ethereum in
Section~\ref{sec:empirical_speedup}.

\subsection{Single-Transaction Concurrency}
\label{sec:cr_speedup}
\cite{saraph2019empirical} proposes a speculative execution technique that works in two phases. In the first
phase, all transactions are executed concurrently, and all transactions that are found to conflict with other
transactions are moved to a sequential `bin'. In the second phase, the transactions in the bin are
executed sequentially. This is done without any a priori knowledge of which transactions cause a conflict, meaning
that the conflicting transactions are executed twice. We derive the following model that captures this
technique. 

Let $\seqtime$ be the execution time of a given block if all of its transaction were to be executed
sequentially. We can assume without loss of generality that the execution of a single transaction takes one
time unit (after all, this is just a scale factor). Let $\ntx$ be the total number of transactions in the
block, so that $\seqtime = \ntx$. Let $\confrate$ be the conflict rate of a block, and $\ncores$ the number of
cores. During the first $\lfloor\ntx/\ncores\rfloor \cdot$ time units of the concurrent phase, all cores are
busy, and $\lfloor\ntx/\ncores\rfloor \cdot \ncores$ transactions can be executed during this phase. The
remaining transactions take a single additional time unit. Hence, the execution time of the first phase takes
$(\lfloor\ntx/\ncores\rfloor+1)$ time units in total. In the second phase, $\confrate \ntx$ transactions need
to be executed sequentially, which takes $\confrate \ntx$ time units. The total execution time of this
protocol, denoted by $\conctime$ is therefore given by $$ \conctime = \lfloor\ntx/\ncores\rfloor+1 + \confrate
\ntx.  $$
To compare the old and new execution times, we define the \emph{speed-up} $\speedup$ in the same way as
\cite{saraph2019empirical}, i.e., as the ratio of the old execution time to the next execution time, or
$\seqtime/\conctime$. For the method described above, the speed-up equals \begin{equation}
\speedup = \frac{\ntx}{\lfloor\ntx/\ncores\rfloor+1 + \confrate \ntx} = \frac{1}{(\lfloor\ntx/\ncores\rfloor+1)/\ntx + \confrate}.
\label{eq:speedup_single}
\end{equation}
If we have perfect prior information about which transactions are going to conflict or not, then we do not need to execute the conflicted transactions twice, leading to even greater potential speed-ups. We assume that obtaining such knowledge requires a numerical pre-processing step that requires $\prepos$ time units. We then only have to process $(1-\confrate) \ntx$ transaction during the first phase. The execution time of this scheme is given by
$$
\conctime = \prepos + \lfloor(1-\confrate)\ntx/\ncores\rfloor+1 + \confrate \ntx.
$$
and a speed-up of
$$
\speedup = \frac{1}{(\prepos + \lfloor(1-\confrate)\ntx/\ncores\rfloor+1)/\ntx + \confrate}.
$$
This leads to large improvements when the conflict ratio is high and when the duration of the pre-processing is small compared to the total execution time. However, in \cite{saraph2019empirical} (Section 5.5) perfect knowledge of the conflicting transactions was not found to have a considerable impact in practice. We note that a further mild improvement is still possible in this case if $\lfloor\ntx/\ncores\rfloor < \ntx/\ncores$, because not all cores will then be busy during the final time unit of the concurrent phase, which means that the sequential phase can be started (and completed) one time unit earlier.

As an example, we consider the two Ethereum blocks of Figure~\ref{fig:ethereum_tdgs}.
Recall that the conflict rate for the two blocks are $40\%$ and $87.5\%$, respectively. If the completely
speculative approach is applied to the block of Figure~\ref{fig:block_1000007}, then the five transactions
would first be executed concurrently, which can be done in $1$ time unit if $\ncores \geq 5$. However, the
last two transactions would need to be rolled back and executed sequentially, which would take $2$ time units.
Hence, the new execution time is given by $3$ time units, and because the old execution time is $5$ time
units, the speed-up equals $5/3$ or roughly $1.67$. Perfect information about the conflicting transaction only
leads to an improvement in the first phase if $\ncores < 5$, and incurs the additional cost of the
preprocessing step. For the block of Figure~\ref{fig:block_1000124}, nearly all transactions must be executed
twice, and the speed-up is minimal: with 16 or more cores, the first phase takes $1$ time unit, but the
sequential phase takes $14$ time units. This leads to a speed-up of $16/15$ or roughly $1.07$. If between 8
and 15 cores are used, then the first phase takes $2$ units, and the speed-up is therefore equal to $0$. If
fewer than 8 cores are used, then the speed-up becomes smaller than 1, which means that performance becomes
worse.

\subsection{Group Concurrency}
\label{sec:lcc_speedup}

As discussed previously, we can improve on the performance of a fully speculative concurrency technique if we
can perfectly predict which transactions are conflicted. As discussed in Section~\ref{sec:sql_queries}, one
way to determine the set of conflicted transactions in a block is to construct the TDG and use breadth-first
search to determine the connected components -- the conflicted transactions are then those which share a
connected component with at least one other transaction. However, instead of executing the full set of
conflicted transactions sequentially, there is still concurrency between the sets of conflicted transactions
that can be exploited. For example, in the block of Figure~\ref{fig:block_1000124}, transactions 1-14 are 
conflicted, but the set of transactions 1-9 do not conflict
with the set of transactions 10-12, etc. This insight informs our approximations based on the group conflict
rate as given via the relative LCC size, as the size of largest connect component is the largest number of
transactions that need to be executed sequentially.

In a system with $n \rightarrow \infty$, each connected component can be assigned to a single core. The
maximum completion time is then $\abslcc$ time units, where $\abslcc$ denotes the absolute LCC size. Because
the old execution time equals $\ntx$ time units and the new execution time equals $\abslcc$ time units, the
speed-up equals $1/\rellcc$, where $\rellcc = \abslcc/\ntx$ is the group conflict rate. If $\ncores$ is
finite, then it is impossible to complete execution in fewer than $\ntx/\ncores$ time units, because this
corresponds to the situation where all cores are busy during the entire execution process. The speed-up in
this case is precisely equal to $\ncores$. However, since it is still not possible to speed up beyond
$1/\rellcc$, the maximum potential speed-up is bounded from above by \begin{equation}
\speedup = \min(\ncores,1/\rellcc). 
\label{eq:speedup_group}
\end{equation}
To establish a lower bound instead of an upper bound, more information about the complete structure of the connected components is known. Determining the optimal schedule to execute the different connected components on a small number of cores is equivalent to the multiprocessor scheduling problem, which is known to be NP-hard \cite{kasahara1984practical}. In the following, we will assume that, for a sufficiently large number of cores, $\min(\ncores,1/\rellcc)$ forms a reasonable approximation of the speedup and the leave the evaluation of this in practice to future work. Also note that a computational step is presumably necessary in this setting, which means that the true speedup is only 
$$
\min\left(\frac{\ntx}{\ntx/\ncores + \prepos},\frac{\ntx}{\ntx/\rellcc + \prepos}\right),
$$
but the difference is negligible if $\prepos$ is small compared to the product of the number of transactions and the execution time per transaction.

\subsection{Potential Speed-Up}
\label{sec:empirical_speedup}

\begin{figure}
    \centering
    \subfloat[][Single-transaction concurrency speed-ups.]{
    	\includegraphics[width=0.4\textwidth]{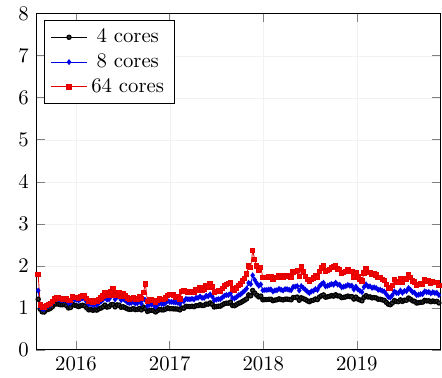}
		\label{fig:eth_single_speedups.pdf}
	} \\
	\subfloat[][Group concurrency speed-ups.]{
    	\includegraphics[width=0.4\textwidth]{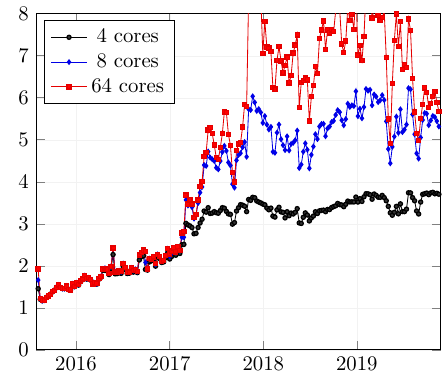}
		\label{fig:eth_group_speedups.pdf}
	}
    \caption{Potential speed-ups for Ethereum, based on single-transaction and group concurrency, respectively.}
    \label{fig:sppedups}
\end{figure}

The approximate speed-ups for Ethereum over its history are shown in Figure~\ref{fig:sppedups} for
different numbers of cores. To construct this graph, we combined \eqref{eq:speedup_single} with the data
of Figure~\ref{fig:ethereumTxConflRatioWeighted.pdf}. It  can be seen in Figure~\ref{fig:eth_single_speedups.pdf} that the speed-ups predicted by the
single-transaction conflict rate are modest, between $1\times$ and $2\times$ depending on the number of cores.
In some cases, the speedup was even lower than $1\times$, which means worse performance than
fully sequential execution. Figure~\ref{fig:eth_group_speedups.pdf}, on the other hand, shows the predictions
made using the group conflict rate, which combined \eqref{eq:speedup_group} with
Figure~\ref{fig:ethereumLccSizeRatioWeighted.pdf}. It can be seen that the speedup in this case is much
higher, up to $6\times$ with $8$ cores and $8\times$ with $64$ cores. We note that to be able to exploit group
concurrency to its full potential, knowledge of the TDG is needed. However, the TDG uses information about
internal transactions that is not available a priori. Nevertheless, an approximate TDG can be constructed by only
using information about the regular transactions. Quantifying the effectiveness of such an approach is left to
future work.

\section{Related Work}
\label{sec:related}
Our work is not the first to look at concurrency in blockchains. \cite{sergey17} shows that there is
inter-block concurrency in Ethereum that arises when a contract {\em communicates} with the external world,
for example, to wait for an input form outside the blockchain. However, its goal is to demonstrate safety
  violations caused by concurrency, whereas our goal is to exploit concurrency for performance.
  \cite{dickerson2017adding} proposes a speculative execution scheme for transactions within the same block. It relies on
  software transaction memory to detect conflicts and perform execution rollbacks. This work is orthogonal to
  ours: in the paper, some concurrency is simulated in a block to validate the proposed technique, whereas our work sheds light on
  how much concurrency there are in existing blockchains.  
  
The work that is most related to ours is \cite{saraph2019empirical}. It also examines potential speed-ups from
exploiting concurrency. It proposes a two-phase technique to do so. First, it executes the
transactions speculatively and detects conflicts at the storage layer. Next, any conflicting transactions are
re-executed sequentially. Our goal is only to quantify concurrency, not the actual execution of transactions,
therefore our approach is more lightweight and lets us analyze more blockchains and more complete data. We are
also able to extract more concurrency than what reported in~\cite{saraph2019empirical}, due to group
concurrency which is not visible at the storage layer. 


In the permissioned setting, \cite{sharma19} examines concurrency in Hyperledger Fabric, which uses a trusted
ordering service instead of a Byzantine fault-tolerant consensus protocol. However, Hyperledger Fabric's
execution is tightly coupled to the ordering service, and it is distinctively different to that of the seven
blockchains we consider. By default, transaction execution in Hyperledger Fabric is concurrent and
speculative. While our goal is to examine how much untapped concurrency there is, \cite{sharma19} is concerned
with a concurrency control mechanism that limits transaction aborts.

\section{Conclusions \& Discussion}
\label{sec:conclusions}
We have presented our analysis of how much concurrency is available in existing blockchains. We have considered two
concurrency metrics: the single-transaction conflict rate, and the group conflict rate. We have examined historical data
from seven public blockchains, and discussed several findings. One finding is that there is more
concurrency in UTXO-based blockchains than in account-based ones, although the amount of concurrency in the
former is lower than expected. Another is that some blockchains with larger blocks have more
concurrency than blockchains with small blocks. Finally, we have proposed an analytical model for estimating
execution speed-up given an amount of concurrency. The model estimates up to 6$\times$ speed-ups in Ethereum
using 8 cores. 

Our work provides insights into a largely unexplored avenue for increasing blockchain performance. However, it
has several limitations that we plan to address in future work. One major limitation is that we have not designed and
implemented an execution engine that can exploit the available concurrency. The main challenge is to minimize
overhead in building the TDG and in scheduling concurrent execution. Another limitation is that we only focused on
inter-transaction concurrency at block level, which leaves other sources of concurrency such as 
intra-transaction, inter-block and inter-blockchain unexplored. Exploiting multiple sources is likely to bring more
performance gains.

\section*{Acknowledgements}
This research/project is supported by the National Research Foundation (NRF), Prime Minister's Office, Singapore, under its National Cybersecurity R\&D Programme and administered by the National Satellite of Excellence in Design Science and Technology for Secure Critical Infrastructure, Award No. NSoE DeST-SCI2019-0009.


\bibliographystyle{abbrv}
\bibliography{ref}


\end{document}